\documentstyle[aps]{revtex}
\begin{document}
\title{Torsional Topological Invariants\\ (and their relevance for real life)\footnote{Lecture presented at the {\em Meeting on Trends in Theoretical Physics} held at La Plata, April 28-May 6, 1997.}} 

\author{Osvaldo Chand\'{\i}a$^{a,b}$ and Jorge Zanelli$^{a,c}$} 

\address{$^{(a)}$ Centro de Estudios Cient\'{\i}ficos de Santiago, Casilla 16443, Santiago, Chile\\ 
$^{(b)}$ Departamento de F\'{\i}sica, Facultad de Ciencias, Universidad de Chile, Casilla 653, Santiago, Chile\\
$^{(c)}$ Departamento de F\'{\i}sica, Universidad de Santiago de Chile, Casilla 307, Santiago 2, Chile.} 

\maketitle
\begin{abstract}
The existence of topological invariants analogous to Chern/Pontryagin classes for a standard $SO(D)$ or $SU(N)$ connection, but constructed out of the torsion tensor, is discussed. These invariants exhibit many of the features of the Chern/Pontryagin invariants: they can be expressed as integrals over the manifold of local densities and take integer values on compact spaces without boundary; their spectrum is determined by the homotopy groups $\pi_{D-1}(SO(D))$ and  $\pi_{D-1}(SO(D+1))$.  

These invariants are not solely determined by the connection bundle but depend also on the bundle of local orthonormal frames on the tangent space of the manifold. It is shown that in spacetimes with nonvanishing torsion there can occur topologically stable configurations associated with the frame bundle which are independent of the curvature.   

Explicit examples of topologically stable configurations carrying nonvanishing instanton number in four and eight dimensions are given, and they can be conjectured to exist in dimension $4k$.  It is also shown that the chiral anomaly in a spacetime with torsion receives a contribution proportional to this instanton number and hence, chiral theories in $4k$-dimensional spacetimes with torsion are potentially anomalous.
\end{abstract}

\section{Introduction}

Despite many years of research and a host of scattered and suggestive results, torsion has remained a curiosity in differential geometry which seems to have no consequences for the real world. General Relativity is perfectly consistent and enormously successful inspite of the seemingly unnecessary restriction, from the theoretical point of view, of assuming that our spacetime has vanishing torsion.  In fact there are very few effects that could in principle be observed in a spacetime with torsion (Einstein-Cartan gravity). Classical particles of different spins would follow different geodesics, but it does not seem to be an easy task to measure that effect experimentally.

A generic Riemannian manifold is endowed with two fundamental and independent entities: a metric and an affine structure. In the traditional approach to gravitation theory, these two notions are linked by postulating the affine connection to be given by the Christoffel symbol and is therefore a function of the metric.  In this way, all local geometric features of the manifold are reflected in the Riemann curvature tensor of the manifold.  If the affine connection is not assumed to be a function of the metric, then the local geometry is endowed with two independent tensors, curvature and torsion. 

It was Elie Cartan who first pointed out the arbitrariness of assuming an affine connection given purely as a function of the metric in the context of Einstein's General Relativity.  It was Einstein's point of view --as well as that of most early researchers in the field-- that torsion would be an unnecessary addition which wasn't required for the most economical and successful theory of spacetime: General Relativity was complicated enough as it was to entertain such extravagances. Cartan, on the other hand, refused to accept that one should limit one's scope to the poor astronomical data available in the 1920's, especially if one accepts with him that the two notions are logically independent and therefore Einstein's proposal was a particular case.

The dialogue between these two stubborn masters is beautifully contained in the letters they exchanged between 1929 and 1932 \cite{C-E}.  Needless to say, it was Einstein's view that prevailed. Thus, in most common texts on gravitation today torsion is set to zero from the start, as there seems to be no compelling experimental reason to relax this condition.  

In recent years, however, a more abstract geometrical approach has been prompted by two incumbent issues: the failure to produce a consistent quantum theory of gravitation starting from Einstein's General Relativity, and the tremendous advances and expectations brought about by string theory and its offsprings: membranes.  The questions facing us are not expected to relate directly to observation. On the contrary, the point is to produce an internally coherent picture.  It is perhaps in this atmosphere ushered by string theorists at the end of this millenium, that exotic ideas such as those of E. Cartan don't seem strange anymore.

The present article is an expanded --and hopefully, improved-- version of our recent article \cite{CZ}. Here more emphasis is given to the calculation of the higher-dmensional torsional invariants and the chiral anomaly, which are superficially discussed in that reference.

\section{Einstein-Cartan spaces} 

Consider a manifold whose metric and affine properties are independently described by two dynamically independent fields: the spin connection, ${\omega^a}_b$, and the local frames (vielbein), $e^a$, respectively \cite{zum}.  As we mentioned above, in the tradition of General Relativity these two fields are assumed to be linked by the torsion-free condition $T^a =0$, where
\begin{equation}
T^a = d e^a + {\omega^a}_b \mbox{\tiny $\wedge$} e^b,
\label{torsion}
\end{equation}
is the torsion two-form related to the torsion tensor by $T^a_{\mu \nu} = e^a_{\lambda} T^{\lambda}_{\mu \nu}$. The expression (\ref{torsion}) is similar to that of the curvature 2-form,
\begin{equation}
{R^a}_b = {d \omega^a}_b +{\omega^a}_c \mbox{\tiny $\wedge$} {\omega^c}_b.
\label{curvature}
\end{equation}

From these definitions, curvature seems to be more fundamental as it depends on the existence of the connection field alone, whereas torsion depends on both the connection and the vielbein.  On the other hand, since on any smooth metric manifold a local frame (vielbein) is always defined, torsion can exist even if the connection vanishes. This implies that in a geometric theory of spacetime the local frame
structure is as basic a notion as the connection and, therefore, torsion and curvature should be treated on a similar footing. 

One can look at curvature and torsion from the point of view of the group of local parallel translations on the manifold.  The generator for parallel transport is, by definition, the covariant derivative $\nabla_{\mu}$ and both torsion and curvature appear in the commutator algebra of these generators in a coordinate basis \cite{lovelock},
\begin{equation}
[\nabla_{\mu}, \nabla_{\nu}] = -T^\lambda_{\mu \nu} \nabla_{\lambda}+ \Re _{\mu \nu},
\label{commutator}
\end{equation}
where $\Re $ stands for the curvature tensor in the same representation as $\nabla_{\mu}$. Here curvature and torsion play quite different roles: $T$ is the structure function for the group of parallel translations, while  $\Re$ is a ``central charge" in the Lie algebra. From this expression it is clear that one can consider equally well two extreme cases: spaces with curvature and no torsion (standard Einstein gravity), and spaces with zero curvature and nonvanishing torsion (``teleparallelizable" geometries) \cite{schouten}. Both possibilities are just special cases of the generic situation.

\section{Topological Invariants: Chern and Euler classes} 

Curvature plays an important role in the characterization of the global topology of a manifold. The integral of the second Chern class --also known as the Pontryagin number-- and the Euler number, defined by
\begin{equation}
P_4 = \frac{1}{8\pi^2}\int_{M_4}R^{ab}{\mbox{\tiny $\wedge$}}R_{ab},
\label{p4}
\end{equation}
and
\begin{equation}
E_4 = \frac{1}{2(4\pi)^2} \int_{M_4}\epsilon_{abdc} R^{ab}\mbox{\tiny
$\wedge$}R^{cd},
\label{e4}
\end{equation}
are well known examples of topological invariants in four dimensions. 

The remarkable feature of these expressions is that, although they are defined purely in terms of local functions, they only take integer values. Now, $P_4$ and $E_4$ are continuous functionals of the curvature and therefore they do not change under continuous deformations of the geometry of $M$. Thus, these characteristic classes label topologically distinct four-geometries. The proof of invariance of expressions like $P_4$ and $E_4$ can be found, for example, in \cite{EGH,nakahara}. 

The Pontryagin number, $P_4$ --or the second Chern class, $C_2$--, counts the difference between the number of selfdual and anti-selfdual harmonic conections on $M$. The Euler characteristic, in turn, equals the alternating sum $\beta_0 - \beta_1 + \beta_2 - \beta_3$, where the Betti number $\beta_p$ stands for the number of harmonic $p$-forms on $M$.

The $n$-th integral Chern class, $C_n$, can be defined for any compact gauge group $G$ and is a topological invariant in a compact $2n$-dimensional compact manifold given by
\begin{equation}
C_{n}[G]= \frac{1}{2} \int_{M_{2n}} c_{n}(A),
\label{Cn}
\end{equation}
where the $2n$-form
\begin{equation}
c_n(A) =Tr\left( \frac{F}{2\pi} \right)^n,
\label{cn} 
\end{equation}
is the $n$-th Chern class for the group $G$.  Here $F = dA + A\mbox{\tiny $\wedge$} A$ is the curvature 2-form for the group $G$ whose generators are normalized as $Tr (G_a G_b ) = \delta_{ab}$, and the wedge product is understood (see \cite{nakahara}). These elementary 2n-forms are the building blocks of the Pontryagin classes, which are costructed by taking linear combinations of all possible products of the form $c_{n_1} c_{n_2} \cdots $, with $2(n_1 +n_2 + \cdots) = D$.

The curvature 2-form for the manifold ($R^{ab}$) is a second rank antisymmetric tensor in the $SO(D)$ indices and therefore the Chern classes for the group of rotations on the tangent space, must contain an even number of curvatures ($n=2k$) and therefore the Chern and Pontryagin classes {\em of the manifold} are only defined in dimensions $D=4k$.

The Euler form, on the other hand,  exists in any even dimension $D=2n$,
\begin{equation}
E_{2n} = \frac{1}{(4\pi)^n n!}\int_M \epsilon_{a_1 \cdots a_{2n}} R^{a_1 a_2}\mbox{\tiny $\wedge$} \cdots R^{a_{2n-1} a_{2n}},
\label{Euler}
\end{equation}
but it cannot be defined for a generic gauge group.

How does one know that the Chern and Euler numbers are topological invariants? The proof goes as follows (see, e.g., \cite{EGH,nakahara}): suppose $K$ is the integral of a $D$-form $k(\omega)$, which is a local function of the connection an its derivatives. Then, if for the deformation $\omega \rightarrow \omega '$, $\Delta = k(\omega ') - k(\omega)$ is the exterior derivative of a globally defined $(D-1)$-form, then the integral $K = \int k$ changes by a surface term under the deformation. Now, if the manifold has no boundary (or the deformation is the identity at the boundary), $K$ would remain unchanged. A physicist's proof would consist in showing that the first order variation of $K$ under infinitessimal $\delta \omega^a_b$ vanishes for a manifold without boundary (or with fixed boundary conditions for $\omega$). Since we are only interested in continuous deformations which can be attained by repeated infinitessimal ones, this proof should be sufficient.

An alternative proof consists in showing that the $D$-form $k(\omega)$ is itself the exterior derivative of something else (the Chern-Simons ($D-1$)-form), $k = d{\cal C}$. Since the functonal variation and the derivative commute, $\delta k = d\delta {\cal C}$, and the rest of the proof is as before.

\section{Torsional Topological Invariants: The Nieh-Yan class} 

Invariants analogous to the Chern classes, but constructed using the torsion tensor are less known. The simplest example of a topological invariant of this sort is provided by the Nieh-Yan four-form (N-Y)\cite{NY}, 
\begin{equation}
N_1 = \int_M n_1,
\label{NY}
\end{equation}
where
\begin{equation}
n_1 = T^a \mbox{\tiny $\wedge$} T_a - R_{ab}\mbox{\tiny $\wedge$} e^a\mbox{\tiny $\wedge$} e^b.
\label{nY}
\end{equation}
This last expression is the only nontrivial locally exact 4-form which vanishes if $T^a =0$ regardless of the curvature, which implies that $n_1$ must be independent of the Chern and Euler classes. In any local patch where the vielbein is well defined, $n_1$ can be written as 
\begin{equation}
n_1 = d(e^a \mbox{\tiny $\wedge$}T_a),
\label{eT}
\end{equation}
and is therefore (locally) exact.  

The topological invariants discussed in the previous section are functionals of the connection and completely independent from the local frame field. In contrast, it is clear from (\ref{eT}) that the N-Y form is a function of the local frame field. 

The invariance of $N_1 = \int_M n_1$ under continuous deformations of $M$ can also be explicitly proven. If $n_1$ and $ n'_1$ are the N-Y densities for ($\omega$, $e$), and $(\omega', e')$, where $\omega' = \omega + \lambda$, $e' =e + \zeta $, then direct computation shows $\Delta= n'_1 - n_1$ is locally exact (a total derivative). Furthermore, if the deformation between $(\omega, e)$ and $(\omega', e')$ is globally continuous, $\Delta$ is globally exact. Therefore, $N$ is a topological invariant in the same sense as the integral Chern and Euler characteristics. Similar invariants can be defined in higher dimensions as will be discussed below. 

The 3-form $e^a\mbox{\tiny $\wedge$} T_a$ is an object that occurs in several seemingly unrelated instances. This Chern-Simons like three form contain first derivatives of the local frames and is therefore a good Lagrangian for the dreibein in three dimensions.  In fact, when added to the standard Chern-Simons Lagrangian, $L_{CS} = \omega \mbox{\tiny $\wedge$} d \omega + (2/3) \omega \mbox{\tiny $\wedge$} \omega \mbox{\tiny $\wedge$} \omega$, it gives rise to the ``exotic" Lagrangian for gravity, which  has the same equations of motion as the Einstein-Hilbert form, $L_{EH} = \epsilon_{abc} (R^{ab} \mbox{\tiny $\wedge$} e^c -(1/3) e^a \mbox{\tiny $\wedge$} e^b \mbox{\tiny $\wedge$} e^c)$. 

The dual of this 3-form  in four dimensions is the totally antisymmetric part of the torsion tensor,
\begin{equation}
e^a\mbox{\tiny $\wedge$} T_a \mbox{\tiny $\wedge$} dx^{\alpha} = \epsilon^{\mu \nu \lambda \alpha} 
T_{\mu \nu \lambda} d^4x.
\label{e-T}
\end{equation}
This component of the torsion tensor, sometimes also referred to as H-Torsion, is the one that couples to the spin 1/2 fields \cite{obukhov}. If the connection is separated into a purely metric part and a torsional part, (\ref{e-T}) is the only conbination that occurs in the Dirac equation. 

Expression (\ref{e-T}) is one of the irreducible pieces of the first Bianchi identity. In a metric-affine space, the
Bianchi identity $D R^A_B = 0$ can be decomposed according to $16 = 3\times 3 + 3 + 3 + 1$. And the  `$1$' is the statement that the Nieh-Yan density is closed, according to (\ref{eT}) \cite{Hehl}.

The Nieh-Yan four-form $n_1$ can also be added as a surface term to the gravitational action in 4 dimensions. This obviously doesn't change the physical contents of the theory and merely produces a canonical transformation. It is this transformation that allows passing from the standard canonical variables of the ADM formalism to the representation in terms of the so-called Ashtekar variables \cite{dolan,bengtsson}.

\section{Relation between the Nieh-Yan and Chern classes}

It seems natural to investigate to what extent the Nieh-Yan invariant (\ref{NY}) is analogous to the Pontryagin and Euler invariants. In particular, it would be intersting to know whether the integral class $N$ over a compact manifold has a discrete spectrum as in the case the of other characteristic classes.  It turns out that there is a natural embedding of the group of rotations in the tangent space, $SO(4)$, into $SO(5)$ that provides a connection between the Chern and the Nieh-Yan classes. This embedding can also be constructed in $D$-dimensions.

Consider a connection for $SO(5)$, $W^{AB}$, whose components combine the spin connection  and the vierbein for $SO(4)$ in the form \cite{MZ,BTZ,G}
\begin{equation}
W^{AB} = \left[
\begin{array}{cc}
\omega^{ab} & \,\,\frac{1}{l}e^a\\
-\frac{1}{l}e^b & \,\,0
\end{array} \right],
\label{embedding}
\end{equation}
where $a, b =1, 2,\cdots 4$ $A, B =1, 2,\cdots 5$. Note that the constant $l$ with dimensions of length has been introduced in order to match the standard units of the connection (inverse length) and those of the
vierbein (dimensionless).  This is the usual embedding of the Lorentz group into the (anti-) de Sitter group for a spacetime with nonvanishing cosmological constant. Then $l$ is the radius of the universe and $|\Lambda| = l^{-2}$ is the cosmological constant.

The inclusion of the parameter $l$ can be viewed as the introduction of a new length scale in the theory. An alternative --and more accurate-- point of view is that $l$ is brought in in order to cancel from $e$ the length scale present in the spacetime manifold so that the enlarged connection $W$ is insensitive to the changes of scale in the manifold.  This means that if one is interested in constructing the curvature form (field strength) and Lagrangians for the enlarged connection, the physically meaningful field in such a theory is not $e$ but $\bar{e} \equiv e/l$. 

The curvature 2-form constructed from $W^{AB}$ is
\begin{eqnarray}
F^{AB} & = & dW^{AB} +W^{AC}\mbox{\tiny $\wedge$} W^{CB}\nonumber \\
	   & = & \left[ \begin{array}{cc}
		R^{ab} - \frac{1}{l^2}e^a \mbox{\tiny $\wedge$} e^b &
\frac{1}{l}T^a \\
			-\frac{1}{l}T^b                         & 0
		\end{array} \right].
\label{F}
\end{eqnarray}

It is direct to check that the second Chern class for $SO(5)$ is the sum of the Chern class for $SO(4)$ and the Nieh-Yan form $n_1$,
\begin{equation}
F^{AB} \mbox{\tiny $\wedge$} F_{AB} = R^{ab}\mbox{\tiny $\wedge$}
R_{ab} + \frac{2}{l^2}[T^a\mbox{\tiny $\wedge$} T_a -
R^{ab}\mbox{\tiny $\wedge$} e_a \mbox{\tiny $\wedge$} e_b].
\label{FF}
\end{equation}
This shows, in particular, that 
\begin{equation}
\frac{1}{2\pi l^2} N_1 = C_2[SO(5)] - C_2[SO(4)].
\label{NCC}
\end{equation}
Hence, it is clear that $N_1$ is indeed a topological invariant as it is the difference of two integral Chern classes. 

As is well known, the n-th Chern class takes on integer values (the instanton number) of the corresponding homoptopy group, $\pi_{2n-1}(G)$ (see, e.g., \cite{nakahara}). In the case at hand, $\pi_3(SO(5)) =$ {\bf Z} and  $\pi_3(SO(4)) =$ {\bf Z + Z}. Hence, from (\ref{NCC}) one can directly read off the spectrum of $N_1$: The integral of the Nieh-Yan form over a compact manifold $M$ must be a sum of three integers. Combining (\ref{Cn},\ref{cn}) and (\ref{NCC}), one finds
\begin{equation}
\frac{1}{(2\pi l)^2}\int_M n_1 = (z_1 + z_2 + z_3), \;\;\; z_i \in
\mbox{{\bf Z}}.
\label{spectrum}
\end{equation}

\section{Torsional Instanton} 
\subsection{The I$\!$R$^4$-Instanton}  

We now show how a geometry with nonvanishing $N_1$ can be easily constructed using the fact that the N-Y form may be nonzero even if the curvature vanishes.  The simplest case occurs in I$\!$R$^4$, where the connection $\omega^{ab}$ can be chosen to vanish identically everywhere. 

Consider now a vierbein field that approaches a regular configuration as $r \rightarrow \infty$. The question is then how to cover the three-sphere at infinity ($S^3_{\infty}$) with a set of 4 independent, everywhere regular  one-forms $e^a$. One possibility is to take one of the $e$' to point along the radial direction, while the other three lie tangent to the $S^3_{\infty}$. But, is it possible to cover the sphere with three linearly independent, globally defined vector fields? The answer is yes.

It is a classical result known as Adam's theorem which proves that $S^1$, $S^3$ and $S^7$ are the only parallelizable spheres, i.e. those which possess a globally defined basis of continuous vector fields \cite{adams}. Defining the sphere through its embedding in I$\!$R$^4$, $x^2 + y^2 + z^2 +u^2 =r^2$, we chose on its surface

\begin{eqnarray}
e^r/l &=&  \frac{1}{r^2}( xdx + ydy + zdz + udu) = \frac{dr}{r} \nonumber \\
e^1/l &=&  \frac{1}{r^2}( ydx - xdy - udz + zdu) \nonumber \\
e^2/l &=&  \frac{1}{r^2}(-zdx - udy + xdz + ydu) \label{dreibein}\\
e^3/l &=&  \frac{1}{r^2}( udx - zdy + ydz - xdu). \nonumber 
\end{eqnarray}

These fields are well defined for $r\neq 0$ and can be smoothly continued inside the sphere (e.g., rescaling them by a function that vanishes as $r\rightarrow 0$ and approaches 1 for $r\rightarrow \infty$). In any case, it is clearly impossible to ``comb" the $S^3_{\infty}$  without producing a singular point where $e^a$ vanishes at finite $r$.  Integrating of $e_a{\mbox{\tiny $\wedge$}}T^a$ over a sphere of radius $r$, one finds\footnote{Strictly speaking, it is not necessary to take the limit $r \rightarrow \infty$. Expression (\ref{int}) is finite and independent of the radius for any finite $r$.}

\begin{equation}
\frac{1}{(2\pi l)^2}\int_{S^3_{\infty}} e_a {\mbox{\tiny $\wedge$}} de^a = 3.
\label{int}
\end{equation}

Thus, using (\ref{eT}), one concludes that the above result is equal to the integral of the Nieh-Yan form over I$\!$R$^4$. Since the integral of $e^a \mbox{\tiny $\wedge$}T_a$ on any sphere gives the same value, one concludes that for the configuration (\ref{dreibein}),
\begin{equation}
n_1 = 3(2\pi l)^2 \delta^{(4)}(\mbox{{\bf x}}).
\label{delta}
\end{equation}
The factor 3 reflects the fact that there are three independent fields summed in the integrand of (\ref{int}).  Configurations with other instanton numbers can be easily generated by simply choosing different winding numbers for each of the three tangent vectors $e^i$. In the example above each of these vectors makes a complete turn around the equatorial lines defined by the planes $x=y=0$, $x=z=0$, and $x=u=0$, respectively. Thus, if the fields $e^i$ make $z^i$ turns around the equators, we recover the result (\ref{spectrum}) in general.

The instanton presented here is very different from the one discussed by D'Auria and Regge also in four dimensions \cite{DR}. That solution is associated to a different singularity in the vierbein structure of the manifold. That solution also has nonvanishing torsion, but has zero N-Y number and nonzero Pontryagin and Euler numbers.

\subsection{The $S^4$ Instanton}  

Consider now $M =S^4 =\{ (x_1, ..., x_5) / (x_1)^2 + \cdots +(x_5)^2 = l^2 \}$ for which
\begin{equation}
R^{ab} = \frac{1}{l^2} e^a {\mbox{\tiny $\wedge$}} e^b.
\label{cc}
\end{equation}
The Chern class $R^{ab}R_{ab}$ vanishes identically because $e^a {\mbox{\tiny $\wedge$}} e_a \equiv 0$, and therefore $c_2[SO(5)] = (2/l^2) n_1$. Thus, the integral N-Y class for different frame configurations on $S^4$ are given by the values of the nontrivial $SO(5)$ connections on $S^4$,
\begin{equation}
\frac{2}{l^2} \int_{S^4} (T^a \mbox{\tiny $\wedge$} T_a - R_{ab}\mbox{\tiny $\wedge$} e^a\mbox{\tiny $\wedge$} e^b) = C_2[SO(5)].
\label{n[s04]}
\end{equation}

In order to evaluate $C_2[SO(5)]$ one can use the Wu-Yang construction \cite{WY}. In fact, it is clear that (\ref{cc}) cannot be a global expression for it is impossible to cover entirely $S^4$ with a blobally defined (everywhere continuous, smooth) vierbein field $e^a_{\mu}$. Therefore, (\ref{cc}) is at best true only on local patches that cover the sphere. This is precisely the reason to use the Wu-Yang construction \cite{pvN}.

Since $tr F^2$ is closed, it can be expressed locally as the exterior derivative of a Chern-Simons three-form $\cal{N}$. One can cover $S^4$ with two patches $S^4_{\pm}$, one on each hemisphere and extending slightly on the other hemisphere, so that the patches overlap on the equator ($x_5=0$). The $SO(5)$ connection on each patch, $W_{\pm}$ are related by a gauge transformation in the overlapping belt around the equator, and therefore one finds
\begin{equation}
\int_{S^4} Tr \left( \frac{F}{2\pi} \right)^2 = \int _{\partial S^4_+} {\cal N}_+ + \int _{\partial S^4_-} {\cal N}_-,
\label{intS4}
\end{equation}
where $\cal{N}_{\pm}$ are the Chern-Simons forms for $W_{\pm}$,  respectively. In the limit when the overlapping belt becomes infinitessimally thin, the r.h.s. of (\ref{intS4}) becomes an integral over $S^3$ of the element of gauge transformation $U$ that that connects $W_+$ and $W_-$ on the equator. Since $U$ is a mapping from $S^3$ onto the gauge group $SO(5)$, this integral takes values that label the third homotopy class of $U$. Since $\pi_3[SO(5)] =$ {\bf Z}, one finds
\begin{equation}
N_1[S^4] = z \in \mbox{\bf Z}.
\label{N1S4}
\end{equation}

This result shows that on $S^4$ there exist configurations of the vierbein field which have integer winding number, although we have not provided an explicit expression for it. These configurations are labeled by one integer, while in the previous example they were labeled by three.

\section{Generalized Nieh-Yan invariants} 

The generalizations of the N-Y form $n_1$ to higher dimensions is straightforward. The number of topological invariants that can be produced in higher dimensions grows wildly and their geometrical interpretation becomes cloudy, their physical consequences are less understood. Consider the elementary forms
\begin{eqnarray}
X_{k} =  e\cdot R^{2k-1} \cdot e   \;\;\;\; (4k)-\mbox{form},\nonumber \\
Y_{k} =  e\cdot R^{2k-2} \cdot T   \;\;\;\; (4k-1)-\mbox{form}, \\
Z_{k} =  T\cdot R^{2k-2} \cdot T   \;\;\;\; (4k)-\mbox{form}, 
\label{c-s}
\end{eqnarray}
where the ($\cdot$) indicates contracted indices. 

It is easy to see that expressions of the form $e\cdot R^{2k} \cdot e$ and $T\cdot R^{2k+1} \cdot T$ vanish identically, while $ \tilde{Y}_l = e\cdot R^{2l+1} \cdot T$ is an exact ($4l+5$)-form ($= -(1/2)d X_l = 1(/2) d Z_l$). It is also direct to show that $X_k$ and $Z_k$ differ by an exact form,
\begin{equation}
d Y_k = Z_k -X_k .
\label{genNY}
\end{equation}

In this notation, the Nieh-Yan four-form is $n_1 \equiv dY_1 =Z_1 - X_1$, and relation (\ref{genNY}) is the generalization of (\ref{eT}). From (\ref{genNY}) it is also clear that the exterior derivatives of $Z_k$ and $X_k$ are the same. 

An exact form of arbitrary degree can be expressed as a product of terms of the form
\begin{equation}
d(X_{k_1} \cdots X_{k_r} Y_{l_1} \cdots Y_{l_s}Z_{m_1} \cdots Z_{m_t}),
\label{monster}
\end{equation}
and linear combinations thereof.  Expression (\ref{monster}) is an exact $(4[k_1 + \cdots + k_r + l_1 +\cdots + l_s +m_1 +\cdots +m_t] -s +1)$-form which is not a product of exact forms of lower degree. Thus, exact forms like this, and their corresponding characteristic classes can be constructed possibly in any dimensions $D\geq 6$. For example, in dimensions 5, 7 and 8, one has
\begin{equation}
d[e\cdot R \cdot e], \;\;\;\;\;\;\;\; D=5
\label{D=5}
\end{equation}
\begin{equation}
d[(e \cdot T)^2],  \;\;\;\;\;\;\;\;\; D=7
\label{D=6}
\end{equation}
\begin{equation}
d[(e \cdot R \cdot e)(e \cdot T)], \;\;\;\; D=8.
\label{D=8}
\end{equation}

There are no torsional topological invariants in six dimensions. This is because there are no Lorentz scalar 5-forms that can be constructed with the torsion. 

It is hard to envissage a classification of all possible torsional invariants in all dimensions. It might even turn out that such a classification is of little or no use and that only a few among all possible invariants of this sort have some relevance in physics. As we shall argue below, however, there are some of these invariants that definitely appear in physical systems, for example as contributions to the chiral anomaly.

Some of the generalized N-Y forms can be related to the higher order Chern classes.  After $D=4$, the next interesting example occurs in eight dimensions, where the relevant Chern class is $c_4 [G] = \frac{1}{2} Tr\left( \frac{F}{2\pi} \right)^4$. Then, the embedding (\ref{embedding}) yields,
\begin{eqnarray}
c_4 [SO(9)] &=& \frac{1}{2} Tr \left( \frac{F}{2\pi} \right)^4 \nonumber \\
	    &=& \frac{1}{2} Tr \left( \frac{R}{2\pi} \right)^4 + \frac{2}{(2\pi)^4l^2}(e\cdot R^3 \cdot e -T\cdot R^2 \cdot T)
+ \frac{1}{(2\pi l)^4} (T\cdot T - e\cdot R\cdot e)^2 - \frac{2}{(2\pi l)^4}[2(e\cdot R\cdot T)(e\cdot T) \nonumber \\
& &+ (e\cdot R \cdot e)^2 - (e\cdot R \cdot e)(T\cdot T)]  \nonumber \\
            &=& c_4[SO(8)] + 4 n_2 +2 (n_1)^2 + 4d(X_1 Y_1).
\label{C4} 
\end{eqnarray}

\subsection{I$\!$R$^8$-Instanton}

Another simple instanton solution exists in $D=8$ and is analogous to the one in I$\!$R$^4$. In eight dimensions there are four independent torsional invariants,
\begin{eqnarray} 
T\cdot R^2\cdot T - e\cdot R^2\cdot e  &=&  n_2  \\
   (T\cdot T -e\cdot R\cdot e)Tr(R^2)  &=&  n_1c_2  \\
        (T\cdot T -e\cdot R\cdot e)^2  &=& (n_1)^2 \\
2(T\cdot R\cdot e)(T\cdot e) + (T\cdot T)^2(e\cdot R\cdot e)  &=&  d(X_1 Y_1).
\label{D8}
\end{eqnarray}
Note that only the third of these invariants, does not vanish for $R^{ab}=0$. Again, one can use fact that the $S^7$ is parallelizable in order to construct the corresponding instanton, essentially repeating the steps of the four-dimensional construction.

Here there is also a relation between the Chern classes for $SO(9)$, $SO(8)$, and the invariants above. The relation is
\begin{equation}
Tr\left(\frac{F}{2\pi}\right)^4 = Tr\left( \frac{R}{2\pi} \right)^4 + 2(T\cdot T)^2 -2(e\cdot R\cdot e)^2 + 8(T\cdot R\cdot e)(e\cdot T)+ 4(e\cdot R^3 \cdot e - T\cdot R^2 \cdot T), 
\label{F4} 
\end{equation}
or, 
\begin{equation}
c_4[SO(9)] = c_4[SO(8)] + 4n_2 + 2(n_1)^2  + d[X_1 Y_1].
\label{c4}
\end{equation}

We note again that the only torsional invariant that survives in an eight-dimensional curvature-free space is $(T\cdot T)^2 = d[(e\cdot T)(T\cdot T)]$. The integration over a seven sphere $x_1^2 + \cdots + x_8^2 =r^2$ embedded on I$\!$R$^8$ can be easily performed using a frame formed by one radial 1-form ($e^r$) and seven orthonormal fields ($e^i$), tangent to $S^7$. 

In order to give an explicit representation of the tangent vectors, we note that each of the tangent vectors $e^i$ on $S^3$ can be generated multiplying the ``radial vector" {\bf x$^r$}$= x_0 + \sigma^1 x_1 +\sigma^2 x_2 +\sigma^3 x_3$, by $\sigma_1$, $\sigma_2$ and $\sigma_3$, correspondingly. In the case of $S^7$, the $e^i$'s are generated using the canonical isomorphism between I$\!$R$^8$ and the octonion algebra in the following manner: Consider the element {\bf x$^r$} of the octonion algebra with components $x_1, \cdots, x_8$ in the canonical basis,
\begin{equation}
\mbox{{\bf x$^r$}} = x_0\tau^0 + x_1 \tau^1 + \cdots x_7 \tau^7, \;\;\; x_i \in  \mbox{I$\!$R}^8,
\label{x}
\end{equation}
where $\tau^0 = 1$, $(\tau^a)^2 = -1, \;\; a=1,...7$ and the products of two $\tau$'s are antisymmetric and satisfy the cyclic symmetry of the Pauli matrices, 
\begin{eqnarray}
\tau^1 \tau^2 = \tau^3, \;\;\; \tau^1 \tau^4 &=& \tau^5, \;\;\; \tau^1 \tau^6 = \tau^7, \nonumber \\ 
\tau^2 \tau^4 = -\tau^6, \;\;\; \tau^2 \tau^5 &=& \tau^7, \;\;\; \tau^3 \tau^4 = \tau^7, \nonumber \\ 
\tau^3 \tau^5 &=& \tau^6.
\label{oct}
\end{eqnarray}

If one substitutes the basis elements $\tau^a$ by $dx^{a+1}$, one obtains a correspondence {\bf x$^r$}$\leftrightarrow e^r$. Multiplying {\bf x}$^r$ by each of the seven $\tau^i$'s, seven orthonormal fields $e^i$ tangent to the sphere are produced.  The first one is
\begin{equation}
e^1 = -x_2 dx_1 + x_1 dx_2 - x_4 dx^3 + x_3 dx^4 -x_6 dx^5 + x_5 dx^6 - x_8 dx^7 + x_7 dx^8,
\label{S7}
\end{equation}
and the rest are left as an exercise for the reader. 

The integral of $(T\cdot T)^2$ over I$\!$R$^8$ for the instanton configuration is thus a combinatorial factor times the volume of the $S_{\infty}^7$. The result is
\begin{equation}
\frac{1}{(2\pi l)^4} \int_{S^7_{\infty}} (e_a{\mbox{\tiny $\wedge$}} de^a)(de_b{\mbox{\tiny $\wedge$}} de^b)
= 7^2.
\label{S7'}
\end{equation}
Again, for different winding numbers ($z_i$) of the seven $e^i$'s, one would obtain $(z_1 +\cdots +z_7)^2$

The idea of the $S^4$-instanton can be repeated now to show the existence of a N-Y instanton on $S^8$. Now the key is that $\pi_7[SO(9)] =$ {\bf Z} (see, e.g., \cite{james}), which implies that the $S^8$ instantons are also labeled by one integer.

\section{Chiral Anomaly}

It is well known that the existence of anomalies like, for instance, the breaking of chiral symmetry by quantum effects, is related to the topological properties the background where the system is immersed. In particular, for a masless spin one-half field in an external (not necessarily quantized) gauge connection $A$, the anomaly for the conservation law of the chiral current is proportional to the second Chern class for the gauge group,
\begin{equation}
{\partial}_{\mu} \left\langle J_5^{\mu} \right\rangle = \frac{1}{4{\pi^2}} Tr F \mbox{\tiny 
$\wedge$} F.
\label{anomaly}
\end{equation}

The question then naturally arises as to whether the torsional invariants discussed above can produce similar physically observable effects \cite{MZ}. 

Kimura \cite{kimura}, Delbourgo and Salam \cite{DS}, and Eguchi, Freund \cite{EF} and Alvarez-Gaum\'e and Ginsparg \cite{AG} evaluated the quantum violation of the chiral current conservation in a four dimensional Riemannian background {\em without torsion}, finding it proportional to the Pontryagin density of the manifold,

\begin{equation}
{\partial}_{\mu} \left\langle J_5^{\mu} \right\rangle =  \frac{1}{2(2\pi)^2}R^{ab}{\mbox{\tiny $\wedge$}} R_{ab}.
\label{grav}
\end{equation}

This result was also supported by the computation of Alvarez-Gaum\'e and Witten \cite{AW}, of all possible gravitational anomalies and the Atiyah-Singer index for the Dirac operator in a curved background, and the complete study of consistent nonabelian anomalies on arbitrary manifolds by Bonora, Pasti and Tonin \cite{BPT}.

It has been sometimes argued that the presence of torsion should not affect the chiral anomaly (see, e.g., \cite{AG,BPT,WZ,KY,mavromatos}). In fact, it is clear  from Eqs.(\ref{curvature},\ref{p4}) that the Chern class is insensitive to the presence of torsion. This does not prove however, that when the space has torsion the anomaly is given by the Chern class only. 

An equivalent approach to calculate the anomaly is using the Atiyah-Singer index theorem.  Then the anomaly  is seen as the difference between the left- and right-handed zero modes of the Dirac operator. It is fairly clear that this number should not jump under a continuous deformation of the geometry. Therefore the index could not change under adiabatic inclusion of torsion in the connection. However, nothing can be said {\em a priori} about the changes of the index under {\em discontinuous} modifications in the field of local frames, as it might happen if a torsionless, spacetime is replaced by one containing a topologically nontrivial configuration of the type shown in the previous section. 

The integral of an anomaly must be a topological invariant \cite{zumino} and therefore the assertion that torsion cannot affect the anomaly would be true if there were no topological invariants that could be constructed out of the torsion tensor independently of those that exist for the curvature. 

Direct computations of the chiral anomaly in spaces with torsion were first done by Obukhov \cite{obukhov}, and later by Yajima and collaborators \cite{KY,Y}. These authors find a number of torsion-dependent contributions to the anomaly, some of which are not densities of topological invariants.  In a related work, Mavromatos \cite{mavromatos} calculates the Atiyah-Singer index of the Dirac operator in the presence of ``curl-free H-torsion", which in our language translates as the condition $d(e\cdot T)=0$. He finds a contribution which, by virtue of this assumption, drops out. 

In all the cases reported in Refs.\cite{obukhov,KY,Y} --and in \cite{mavromatos} if one doesn't assume $d(e\cdot T)=0$--, the N-Y term appears among many other.  Many of these torsional pieces, including the N-Y term, are divergent when the regulator is removed, which was interpreted as an indication that these terms were regulator artifacts and should therefore be ignored. 

The calculation using the method of Fujikawa \cite{fujikawa,D-R} is straightforward and we include it here in order to clarify some subtle points which were not thoroughly discussed in \cite{CZ}. Consider a massless Dirac spinor on a curved background with torsion. The action is
\begin{equation}
S=\frac{i}{2} \int d^4x e {\bar{\psi}}{\not \!\nabla}\psi + h.c.,
\label{action}
\end{equation}
where the Dirac operator is
\begin{equation}
i{\not \! \nabla} = i{e_a}^{\mu} {\gamma}^a \nabla_{\mu}.
\label{dirac}
\end{equation} 
Here ${e_a}^{\mu}$ is the inverse of the tetrad ${e^a}_{\mu}$, ${\gamma}^a$ are the Dirac gamma matrices and in what follows $\nabla_\mu$ will denote the covariant derivative for the $SO(4)$-connection in the appropriate representation. 

This action is invariant under rigid chiral transformations
\begin{equation}
\psi {\longrightarrow} e^{i\varepsilon {\gamma}_5}\psi,
\label{chiral}
\end{equation}
where $\varepsilon$ is a real constant parameter. This symmetry leads to the classical conservation law 
\begin{equation}
{\partial}_{\mu}J^{\mu}_5 = 0
\label{conservation}
\end{equation}
where $J^{\mu}_5 = e {e_a}^{\mu} {\bar{\psi}} {\gamma}^a {\gamma}_5 \psi$.

The chiral anomaly is given by  
\begin{equation}
{\partial}_{\mu} \left\langle J^{\mu}_5 \right\rangle = {\cal A}(x),
\label{naive}
\end{equation}
where 
\begin{equation}
{\cal A}(x) = 2\sum_{n} e(x) {\psi}_n^{\dagger} {\gamma}_5 {\psi}_n =2 Tr[\gamma_5].
\label{infinite}
\end{equation}

In order to make sense of this formal expression, $\cal A$ is tentatively regularized as 
\begin{equation}
{\cal A}(x) = 2\lim_{y\rightarrow x} \lim_{\beta\rightarrow 0} Tr \left[{\gamma}_5 \exp(\beta{\not \!\nabla}^2)\right] {\delta}(x,y).
\label{reg}
\end{equation}

This is the standard expression for the anomaly used when the torsion vanishes and we provisionally adopt it here for pedagogical reasons. 

The first thing we observe is that the regulator $\beta$ need not be taken to zero in order to regulate the trace (\ref{infinite}). The reason is that for each nonzero eigenvalue of ${\not \! \nabla}^2$, there are two states of opposite chirality and therefore they cancel pairwise in the trace. The only only remaining contribution to (\ref{reg}) comes from the zero modes and on those states the exponential of ${\not \! \nabla}^2$ is just the identity. Thus the anomaly (\ref{reg}) equals the number of right-handed ($\nu_+$) minus the number of left-handed ($\nu_-$) zero modes, or
\begin{equation}
\int {\cal A}(x) = \nu_+ - \nu_-,
\label{nu}
\end{equation}
which is an alternative representation for the anomaly as the index of the Dirac operator \cite{nakahara,zumino}.

The square of the Dirac operator is given by
\begin{equation}
{\not \! \nabla}^2 = \nabla^{\mu}\nabla_{\mu} - e^{\mu}_a e^{\nu}_b e^{\lambda}_c J^{ab} T^c_{\mu \nu}\nabla_{\lambda} + \frac{1}{2} e^{\mu}_a e^{\nu}_b J^{ab}J^{cd} R_{cd \mu \nu},
\end{equation}
where ${J}_{ab}=\frac{1}{4} [{\gamma}_a , {\gamma}_b ]$ is the generator of $SO(4)$ in the spinorial representation.

The Dirac delta on a curved background can be represented as
\begin{equation}
\delta (x,y) = \int \frac{d^4k}{(2\pi )^4} e^
{k^{\mu}{\nabla}_{\mu}{\Sigma}(x,y)},
\label{28}
\end{equation}
where ${\Sigma}(x,y)$ is the geodesic biscalar \cite{dewitt} connecting the points $x$ and $y$.  The integral over the ``wave vector" $k^{\mu}$ requires some careful handling. The pacetime manifold over which (\ref{reg}) is evaluated will be taken to be a compact Euclidean space ({\it e.g.}, $S^4$) with a typical length scale $l$ --``the radius of the universe". (This is ensures that the tangent space symmetry $SO(4)$ can be embedded into $SO(5)$). Thus, $k$ must be quantized in multiples of the inverse radius, $k^{\mu}\sim  2\pi n^{\mu}/l$, with $n^{\mu} \in$ {\bf Z}. Now, since $l$ is supposed to be very large, the wave vectors $k^{\mu}$ can be approximated by  a continuous variable. This means that the integrations over $k$ yield inverse powers of $l$, which we normalize as
\begin{eqnarray}
                \int \frac{d^4k}{(2\pi)^4} &=&  l^{-4}  \nonumber \\
\int \frac{d^4k}{(2\pi)^4} k^{\mu} k^{\nu} &=&  l^{-6} g^{\mu \nu} \nonumber \\
&& \mbox{etc.}
\end{eqnarray}

Applying the operator $\exp(\beta {\not\!{\nabla}}^2)$ on (\ref{28}), taking the limit $y{\rightarrow}x$, and tracing over spinor indices, one finds
\begin{equation}
{\cal A}_{\beta} = \frac{1}{8{\pi}^2}[-2(\beta/l^2)l^{-2}e\cdot R\cdot e + (\beta/l^2)^2 R^{ab}{\mbox{\tiny $\wedge$}}R_{ab} + 2(\beta/l^2)^2 l^{-2} T\cdot T] + O[(\beta/l)^{-2})].
\label{result}
\end{equation}

In the standard calculations (e.g., in \cite{BPT,WZ,KY,mavromatos}), the length scales $l\sim \beta^{1/2}$ are identified with $M^{-1}$. This means that only the second term in (\ref{result}) would be finite, while the first and the third diverge like $M^2$ and the terms $O[(\beta/l)^{-2})] \sim M^{-2}$ are neglected. 

In our case, we see that if one identifies $\beta$ with $l^2$ then the expression for the anomaly is finite to all orders and the first three terms in (\ref{result}) are
\begin{equation}
\frac{1}{8\pi^2} \left[ R^{ab}\mbox{\tiny $\wedge$}R_{ab} + \frac{2}{l^2} (T^a{\mbox{\tiny $\wedge$}}T_a - R_{ab}{\mbox{\tiny $\wedge$}}e^a{\mbox{\tiny $\wedge$}}e^b )\right].
\label{result'}
\end{equation}
which is the Chern class for $SO(5)$ (\ref{FF}).

It is interesting to observe that if the identification $l = \beta^{1/2} = M^{-1} $ is performed in the results of refs. \cite{obukhov,KY,Y}, all but one of the torsional contributions to the anomaly vanish in the limit when the regulator is removed and the only remaining term would have been $N$.

There is a puzzling point about this result. We started out with an $SO(4)$-invariant regulated expression $Tr[\gamma_5$exp$(\beta \not \! \nabla^2)]$, and ended up with the Chern class for $SO(5)$. How was the extra symmetry smuggled into the anomaly? One way to understand this is by observing that if $\beta =l^2$ the expression in the regulator is the square of the ``dimensionless Dirac operator", $\not \! \nabla = l\gamma^a e_a^{\mu} \nabla_{\mu}$. This operator is constructed using the spin connection $\omega^{ab}$ and the rescaled vierbein $\bar{e}^a \equiv l^{-1}e^a$, which are the components of the $SO(5)$ connection [c.f., (\ref{embedding})]. 

The suspicion might arise then as to whether the first few terms of the series shown in the result (\ref{result'}) are all that is produced. This is a theorem one could try to prove by explicit computation, but that is not really necessary.  We know the result must be an exact four form constructed from the only ingredients we have at our disposal: the invariant tensors of $SO(4)$ ($\delta^{ab}$, $\epsilon^{abcd}$), the curvature and torsion two-forms, and the dimensionless vierbein $\bar{e} = e/l$.  It was shown in \cite{MZ} that the only invariants of this type which do not use inverse vielbeins in four dimensions are the Euler form, the second Chern class and the Nieh-Yan form. The inverse vierbein $e^{\mu}_a$ actually enters in the definition of the Dirac operator (\ref{dirac}) and therefore one could not rule out the occurence of other closed four form constructed with it, like 
\begin{equation}
f\mbox{\tiny $\wedge$}f,
\label{f}
\end{equation}
where the 2-form $f= dA$ is an abelian curvature constructed with the torsion tensor and the inverse vierbein, 
\begin{equation}
\tilde{A}_{\mu} \equiv e^{\nu}_a T^a_{\mu \nu}
\label{A}
\end{equation}

The reader can check that (\ref{f}) has the correct dimensions and could therefore also contribute to the anomaly. The possibility of encountering other terms constructed with $\tilde{A}$ in four dimensions is ruled out by the fact that no other closed 4-forms are known in the presence of an electromagnetic field. 

A contribution of the form (\ref{f}) to the anomaly can be traced back to the Dirac operator as well. Instead of the ``minimal" Dirac operator (\ref{dirac}) one could have used its sel-adjoint extension, which turns out to be
\begin{eqnarray}
i{\not \! \nabla}' &=& \frac{i}{2} ({\not \! \nabla} - {\not \! \nabla}^{\dag})\nonumber \\
&=& i({\not \! \nabla} + \frac{1}{2} \tilde{A}).
\label{dirac'}
\end{eqnarray}
Then, there is certainly a contribution of the form (\ref{f})to the anomaly arising from the ``electromagnetic potential" $\tilde{A}$. Clearly, this anomaly can be cancelled if the fermion field is coupled to an external true electromagnetic field.

\section{Discusion}

\subsection{Higher dimensions} 

As we already mentioned, there exist a large collection of topological invariants in higher-dimensional spacetimes with torsion. Obvious families of these invariants for dimensions $D=4k$, are:
\begin{eqnarray}
(&n&_1)^k,\\
&n&_k = d[T\cdot R^{2k-2} \cdot e],
\end{eqnarray}
and, in general 
\begin{equation}
n_{k_1} n_{k_2} \cdots n_{k_p},
\end{equation}
where $k_1 + k_2 +\cdots k_p = k =D/4$.
There are other invariants which do not fall into these classes, as for example expressions of the form (\ref{monster}). The number of independent torsional invariants for a given dimension follows no simple pattern with $D$, but it can be easily seen is that they occur for almost all dimensions beyond 4. An exceptional case is $D=6$, where no torsional invariants can be defined. 

Does this mean that one should expect to find instantons and anomalies in most dimensions? The answer is not known, but one can conjecture that these invariants only contribute to anomalies in dimensions $D=4k$. The reason is that, by a general argument (see, e.g., \cite{nakahara}), the anomaly is always given by the ($D/2$)-th Chern class of the gauge group.  In our case the relevant gauge groups are $SO(D+1)$ and $SO(D)$. The corresponding Chern classes are nonzero if $D/2$ is even, (i.e., $D=4k$). 

In analogy with the cases $D=4, 8$, one could expect to find torsional contributions to the anomaly, provided both $\pi_{D-1} [SO(D+1)]$ and $\pi_{D-1} [SO(D)]$ are nonzero and not equal to each other. This happens for $D=4, 8, 12, 16$, and seems to be a general feature of $D=4k$ \cite{james}. In all these cases the $S^D$ instantons could exist and they would be labeled by a single integer.

Note that the trick for constructing the I$\!$R$^4$ and I$\!$R$^8$ instantons cannot be repeated in other dimensions because $S^1$, $S^3$ and $S^7$ are the only spheres which admit a globally defined basis of tangent vector fields \cite{adams}. In general, the maximum number of independent global vectors that can be defined on $S^{n-1}$ is given by Radon's formula \cite{husemoller},
\begin{equation}
\rho_n = 2^c +8d -1,
\label{rho}
\end{equation} 
where $n$ is written as $n = [\mbox{odd integer} \times 2^c 16^d]$, with $c$  and $d$ positive integers, and $c\leq 3$ . From this formula, it is clear that for all odd-dimensional spheres $\rho_n \geq 1$, while for even-dimensional spheres (odd $n$), $\rho_n =0$.  On the other hand, for $n-1=$ odd, there is always at least one global vector field on $S^{n-1}$. The connection between the number of independent vectors on $S^{n-1}$ and the instanton spectrum will be discussed elsewhere.

\subsection{Anomaly}

In the previous section we argued that the anomaly could be made finite if  the regulator is chosen as $\beta = l^2$. Under this choice, two things happen:

i) The relevant Dirac operator that entered in the regulator could be written as $\bar{e}^{\mu}_a \gamma^a \nabla_{\mu}$, where $\bar{e}^a_{\mu} = l^{-1}e^a_{\mu}$, which is the way the vielbein enters in the embedding $(\omega, e) \rightarrow W^{AB}$. In agreement with this, the anomaly is the second the Chern class for $SO(5)$, instead of being the second Chern class for $SO(4)$ (as in the torsion-free case).

ii) In terms of the ``physical" fields $\omega$ and $\bar{e}$, the regulator $\beta=l^2$ drops out from the trace before the limit $\beta \rightarrow 0$ is performed.  In other words, the result should be correct to all orders in powers of $\beta$. This is because the limit $\beta \rightarrow 0$ in (\ref{reg}) is actually unnecessary: as we mentioned above, the trace erases all $\beta$-dependence. Thus, the expression on the right hand side of (\ref{result'}) should be independent of $\beta$ {\it before the limit is performed}. 

It should be stressed that the choice $\beta =l$ is the only one needed to yield a $\beta$-independent result, and there seems to be no other similarly simple adjustment that would do the trick. For example, if one had chosen $\beta' = \alpha l$, the result would not be an exact form because this would change the relative factor between the two terms in the NY form.

\subsection{Index} 

The alternative form of the anomaly given in (\ref{nu}) is known as the Atiyah-Singer index for the Dirac operator.  It is well known that in the absence of torsion the index is given by the integral of the second Chern class (Pontryagin number). Obviously, as $C_2(SO(4))$ is independent of the vierbein, its invariance  under continuous deformations of the geometry also allows for continuous deformations of the local frames and, in particular, for the addition of torsion. As is shown here, the presence of torsion could affect the index of the Dirac operator through the addition of the N-Y invariants because the relevant Chern class would be $C_2(SO(5))$.

In \cite{mavromatos} it is shown that there is a torsional contribution to the index although it is set equal to zero by the aditional requirement of curl-free H-torsion, and our result (\ref{result'}) does not contradict that conclusion: curl-free H-torsion is identical to the demand that $n_1 =0$. 

\section{Concluding remarks}

What have we learned from this exercise?  Apparently three things:

{\bf A.}   That a $D$-dimensional Riemannian space can be conceived as one in which the usual $SO(D)$ connection and the (appropriately normalized) vielbein are viewed as parts of an $SO(D+1)$ connection. The torsion in this space is just a piece of the curvature two-form for the $SO(D+1)$ connection.

{\bf B.}   That certain closed $D$-forms constructed out of the torsion two-form are topological invariants. With the construction {\bf A} in mind, these invariants can be related to the Chern classess of $SO(D)$ and $SO(D+1)$.

{\bf C.}   That these invariants may contribute to the chiral anomaly and hence, the quantum theory can in principle ``detect" the presence of torsion in space.

The point of view presented here might become important, for instance, when dimensionally reducing a geometrical theory. Some of the components of the curvature tensor in the original spacetime could be interpreted as torsion in the reduced space \cite{bellisai}.  

In the end, torsion may continue to be an obscure object of research whose physical implications are hard to grasp. We hope, however, we have offered a new angle which could help in dispelling the mystery surrounding it.

{\bf Note Added}

In the past few months, two articles have appeared where the anomaly for a space with torsion is evaluated. In the first, by Obukhov, Mielke, Buczies and Hehl \cite{OMBH}, the anomaly is computed using the heat kernel regularization and our result (\ref{result}) is reproduced. In the same article, a slight modification of our instanton solution in I$\!$R$^4$ and I$\!$R$^8$ is presented. Susequently, another article by Mielke and Kreiner \cite{MK} has appeared, where it is claimed that there can be no torsional contributions to the anomaly, contradicting the previous result by one of those same authors. Also, there are two recent articles where the chiral anomaly in the presence of torsion is considered. In Bellisai, in \cite{bellisai}, discusses the anomaly in the context of string-induced field theory, and in \cite{SC}, Soo and Chang examine the possibility of CPT violation arising from torsion in spacetime.

\section*{Acknowledgments}

We thank the organizers of the Meeting on Trends in Theoretical Physics for their wonderful hospitality at La Plata, which even vegetarians could appreciate fully. We have enormously benefitted from many fruitful discussions and inspiring insights from J. Alfaro, M. Ba\~nados, J. Gamboa, A. Gomberoff, E. Gozzi, F. Hehl, C. Mart\'{\i}nez, Y. Obukhov, G.'t Hooft, R. Troncoso and L. F. Urrutia.  We are deeply indebted to R. Baeza, M. S. Narasimhan and G. Thompson for enlightening discussions and for their patience in trying to understand our questions in poor mathematical language. This work  was partially supported by grants 1960229 and 1970151 from FONDECYT-Chile, and grant 4-9531ZI from DICYT-USACH.  We would also like to thank UNESCO and the International Atomic Energy Agency for hospitality at the International Centre for Theoretical Physics, where most of this work was written up. The institutional support by a group of Chilean companies (EMPRESAS CMPC, BUSINESS DESIGN ASS., CGE, CODELCO, COPEC, MINERA ESCONDIDA, NOVAGAS and XEROX-Chile) is also recognized. J.Z. holds a John Simon Guggenheim fellowship.


\end{document}